\documentclass[ amsmath,amssymb,aps,pra,reprint,floatfix]{revtex4-1}

\usepackage{graphicx}
\usepackage{dcolumn}
\usepackage{bm}
\usepackage{textcomp}

\begin{document}

\title{Characterizing errors on qubit operations via iterative randomized benchmarking}
\author{Sarah Sheldon}
\author{Lev S. Bishop}%
\author{Easwar Magesan}%
\author{Stefan Filipp}%
\author{Jerry M. Chow}
\author{Jay M. Gambetta}
\affiliation{IBM T.J. Watson Research Center, Yorktown Heights, NY 10598, USA}

\date{\today}

\newcommand{\ket}[1]{\left|#1\right>}

\begin{abstract}
With improved gate calibrations reducing unitary errors, we achieve a benchmarked single-qubit gate fidelity of $99.95\%$ with superconducting qubits in a circuit quantum electrodynamics system. We present a method for distinguishing between unitary and non-unitary errors in quantum gates by interleaving repetitions of a target gate within a randomized benchmarking sequence. The benchmarking fidelity decays quadratically with the number of interleaved gates for unitary errors but linearly for non-unitary, allowing us to separate systematic coherent errors from decoherent effects.  With this protocol we show that the fidelity of the gates is not limited by unitary errors, but by another drive-activated source of decoherence such as amplitude fluctuations.  
\end{abstract}

\pacs{Valid PACS appear here}
\maketitle


Accurate characterization of control gates is an essential task for developing any quantum computing device. Quantum process tomography (QPT)~\cite{chuang97,poyatos97,chow09} has been the standard method for characterizing quantum
gates because, ideally, it produces a full reconstruction of the quantum process. In practice however, QPT suffers from many drawbacks, the most inimical being its exponential scaling in the number of quantum bits (qubits) comprising the system and that it is limited by state preparation and measurement (SPAM) errors. Various methods such as randomized benchmarking (RB)~\cite{emerson05, knill08,magesan11,martinis} and gate set tomography (GST)~\cite{Merkel13,blumekohout} have recently been developed to help overcome these limitations. RB is both insensitive to SPAM errors and efficient~\cite{magesan12}. However, it only extracts a single piece of information, the average gate fidelity. GST on the other hand helps to overcome limitations from SPAM errors by reconstructing an entire library of gates in a self-consistent manner. The price paid for this self-consistent reconstruction is an even worse scaling than QPT. 

As control calibration techniques continue to improve and quantum
gates approach the fidelity required for fault tolerant quantum computation, it becomes both important and difficult to
verify the presence of increasingly small errors. Error verification constitutes a critical first step in a debugging routine since different physical mechanisms can lead to different error types. QPT and GST are often poor choices for error verification since they are time consuming and contain so much information that backing out the presence of specific error types on small scales can be a challenge in itself. In addition, SPAM errors in QPT sets a lower limit on the detectable error strengths \cite{Merkel13}. At the other end of the spectrum, while standard RB is efficient the information it contains about the gate is typically not enough to perform any sort of useful error verification. An extension of standard RB, interleaved randomized benchmarking, consists of interleaving a target gate in a benchmarking sequence and provides bounds on the error for the gate of interest~\cite{magesanIRB,gaebler}.  
Interleaved benchmarking can identify gates that are poorly calibrated, but does not reveal if the errors are due to decoherence, over-/underrotations, or off-resonance effects amongst other error types. Thus, fast and reliable routines that determine the presence of specific error types are required.  Others have proposed to use RB for measuring the unital part of a quantum map \cite{wallman}, correlated errors on a multi-qubit space \cite{corcoles}, and recently Ref.~\cite{kimmel} has proposed an alternative method for measuring unitary errors.
In this paper we propose and experimentally implement a protocol, largely based on the ideas of RB, that verifies the presence of unitary versus non-unitary errors.

\begin{figure}
		\begin{align} &\mathrm{(a)}\nonumber\\&\includegraphics[width=\columnwidth]{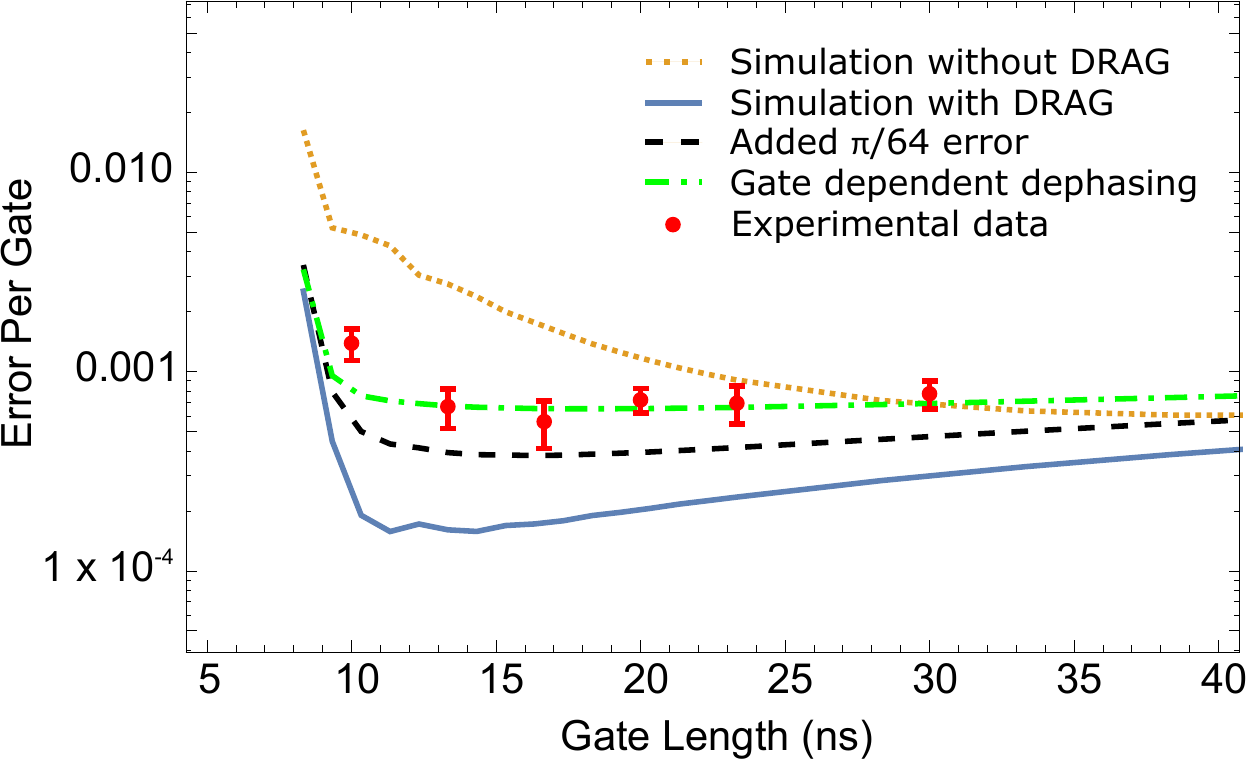}\nonumber\\
		&\mathrm{(b)}\nonumber\\ &\includegraphics[width=\columnwidth]{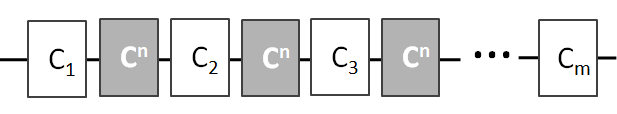} \nonumber \end{align}
		\caption{(color online)(a) Randomized benchmarking fidelity as a function of gate length.  Simulated fidelity with a DRAG correction in solid blue and without in dotted brown. Experimental data (points), with the highest fidelity of 0.9995 occuring at $16.7\,\text{ns}$.  Dashed black line: simulated fidelity when all gates are overrotated by $\pi/64$ (which would be detectable by IRB). Green dot-dashed line: simulated fidelity with gate-dependent dephasing proportional to the drive amplitude $\gamma_\phi=k\Omega$. (b)~The iterative benchmarking sequence with target gate $C$ repeated $n$ times between random Clifford gates, $C_i$. The case $n=0$ corresponds to a regular randomized benchmarking sequence as used for the data in (a).}
	\label{fig:tickPlot}
\end{figure}

A major source of unitary errors in transmon qubits originates from the presence of higher levels, which can be removed by the derivative removal via adiabatic
gate (DRAG) protocol \cite{motzoi}.  To quantify this error source, we compare experimental randomized benchmarking fidelities for several gate times with two simulations, one assuming a DRAG-corrected pulse shape and the other without DRAG (Fig. \ref{fig:tickPlot}).  The measurements described here are performed on a two-qubit sample consisting of two transmon qubits coupled by a coplanar waveguide resonator, with independent readout resonators for each qubit.  The qubit of interest has a transition frequency of $5.0154\,\text{GHz}$ and anharmonicity of $-323\,\text{MHz}$.  $T_1$ and $T_2$ are $45\pm6\,\text{\textmu s}$ and $53\pm10\,\text{\textmu s}$, respectively.  These characteristic times are the mean values from 500 measurements taken over 14 hours, and the error bars are the standard deviation of this data; each independent experiment is well fit by an exponential decay.  The pulses used in the RB sequence are truncated Gaussian pulses having total length equal to four times the standard deviatiation of the Gaussian and with the DRAG correction applied to the quadrature component.

A typical benchmarking sequence consists of a set of random Clifford gates that together compose to an identity operation~\cite{magesan11}. Under realistic assumptions on the noise, the fidelity between the implementation of this sequence with the identity operation decays exponentially as a function of the number of Clifford gates \cite{magesan12}. When the fidelity decay is averaged over many realizations of the random sequence, the decay constant serves as the single metric for the average noise in the system. 

The weak anharmonicity, $\delta$, of the transmon limits the gate fidelity as $1/\delta$, which can be seen for short gate times in Fig. \ref{fig:tickPlot}.  The experimental data falls below the non-DRAG curve (brown dotted line in Fig. \ref{fig:tickPlot}), showing that we have partially removed unitary errors due to presence of higher levels in the transmon. At the gate length $t_g = 16.7\,\text{ns}$, the error rate corresponds to an average fidelity per gate of $99.95\%$ but is not yet limited by $T_1$ and $T_2$ with the DRAG correction (blue solid line). With the current of control, we can calibrate pulses to within a factor of four of the  limit set by $T_1$ and $T_2$, but it is clear that there are still errors remaining in the system.  (The remaining simulations in Fig. \ref{fig:tickPlot} will be described later in this text).  

For longer pulses the fidelity is limited by the finite coherence time of the qubit.  The tradeoff between decoherence and unitary errors shown in Fig.~\ref{fig:tickPlot} is generic across quantum computing hardware. For optimal fidelity, any quantum processor will be operating with fidelity at least partially limited by unitary errors: if this were not the case, then the fidelity could surely be improved by shortening the gate time.

We extend interleaved randomized benchmarking by repeating a target Clifford $n$ times between the random Clifford gates and measuring the fidelity as a function of $n$ repetitions [Fig. \ref{fig:tickPlot}(b)]. If the gate errors are non-unitary, then the fidelity will only depend on the total length of the interleaved segment, and the resulting error per segment will thus be linear with $n$.  If there are unitary errors of an over-/underrotation type, they will add coherently with $n$, and the fidelity  decay will be quadratic to leading order. To see this, suppose we have a single-qubit unitary error of the form
\begin{equation}
U=\exp\left(-i\frac{\epsilon}{2} \hat{r}\cdot \vec{\sigma}\right),
\end{equation}
where $\epsilon$, $\hat{r}$, and $\vec{\sigma}$ are the error angle, axis of rotation, and vector of Pauli operators respectively. Assuming $\epsilon \ll 1$ we can write $U^n$ to second order in $\epsilon$ as
\begin{align}
U^n &= \openone - i n \frac{\epsilon}{2}\hat{r}\cdot \vec{\sigma} - \left(n(2n-1)\right)\frac{\epsilon^2}{4} \left(\hat{r}\cdot \vec{\sigma}\right)^2 + O\left(\epsilon^3\right).
\end{align}
The average fidelity $F$ of the error gate compared to the identity is given by $F= \left(\left|\text{tr}\left(U^n\right)\right|^2 + 2\right)/6$ and writing $F$ in terms of the benchmarking parameter $\alpha = 2F-1$ gives \cite{magesan11}
\begin{align}
\alpha &= 1-\left(\frac{n(2n-1)\epsilon^2}{3}\right),
\end{align}
which shows the quadratic dependence in $n$.  A similar analysis finds that errors due to a $T_1$ or $T_2$ process do decay linearly in $n$.

\begin{figure}
		\includegraphics[width = 0.9\columnwidth]{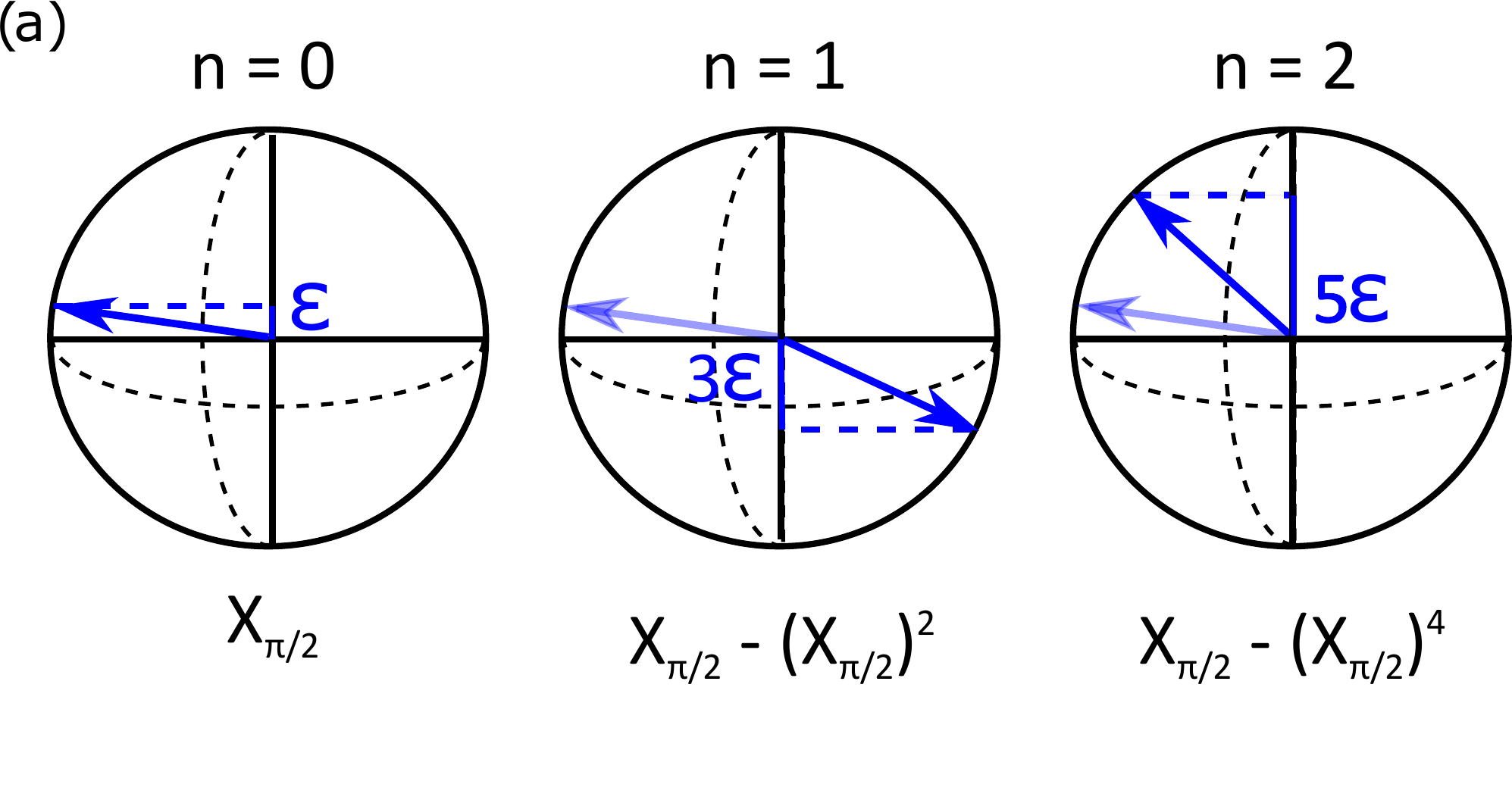}
		\includegraphics[width=0.48\columnwidth]{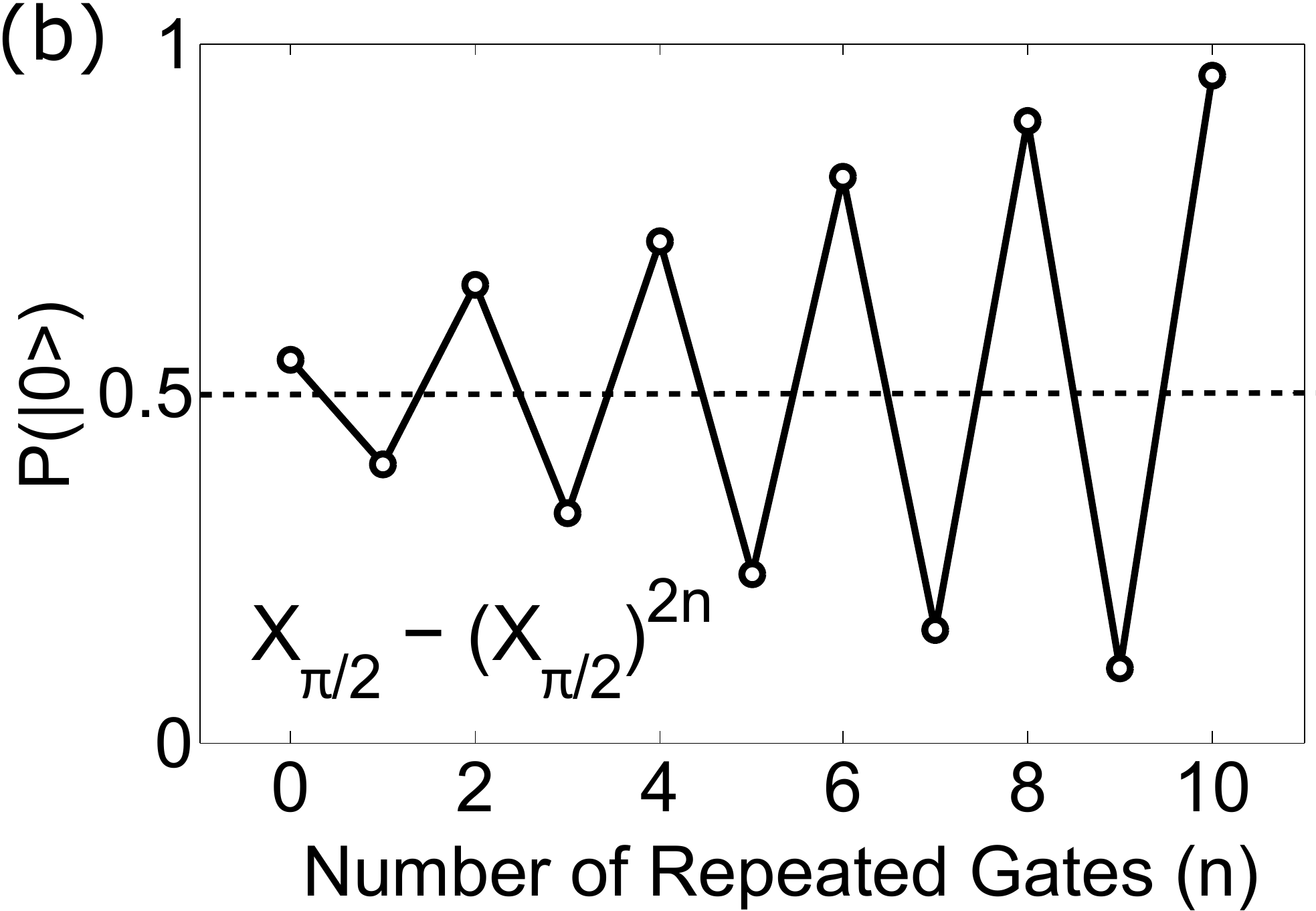}
		\hspace{1mm}\includegraphics[width=0.49\columnwidth]{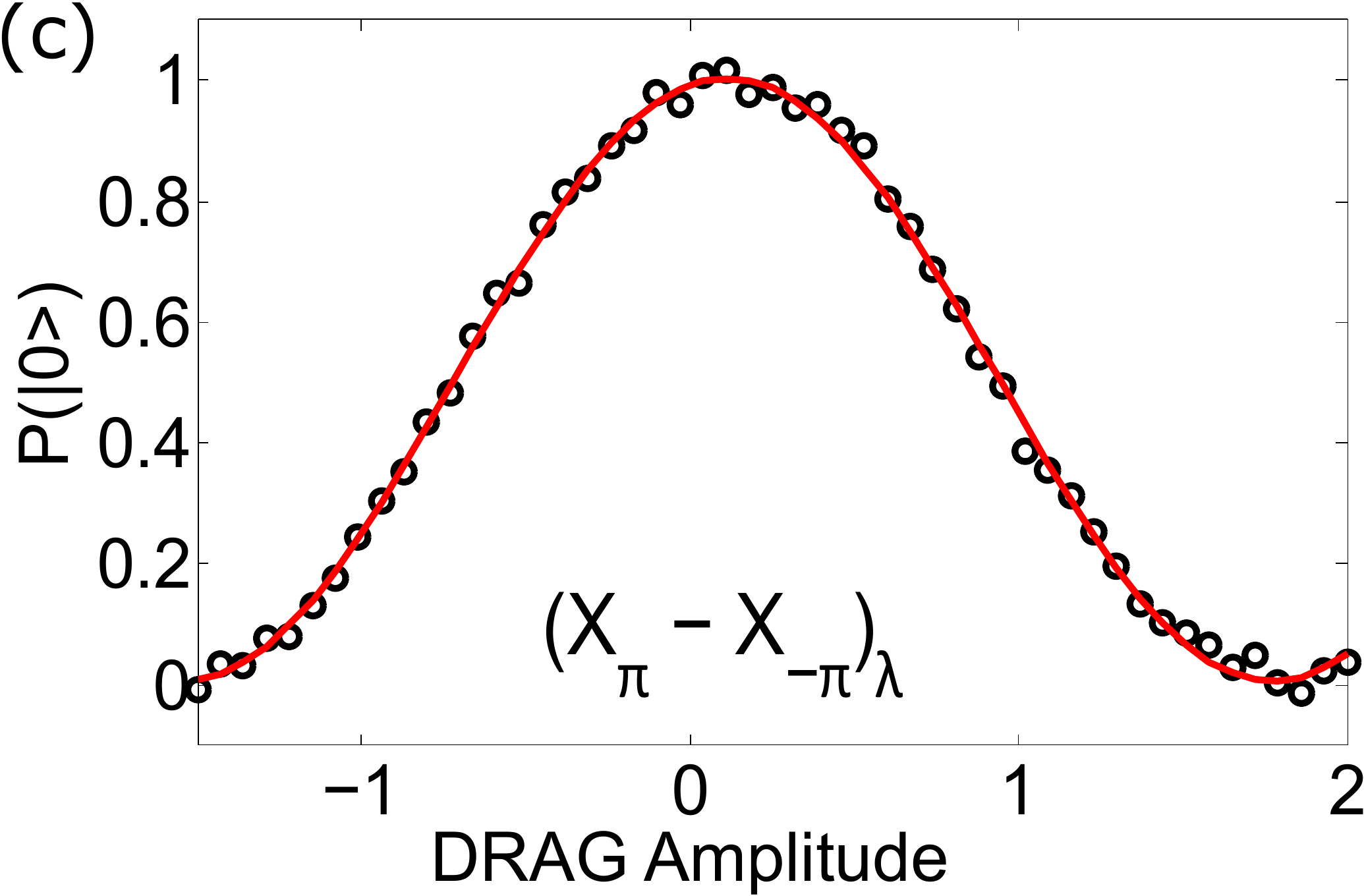}
		\caption{Calibrations of the control pulses: (a) Bloch sphere depiction of the qubit for the first three points of the error amplification sequence given in Eq \label{eq:calFit}. (b) The amplitude calibration for a $X_{\pi/2}$ pulse.  The initial guess for the pulse amplitude has some error, which the sequence amplifies so the deviation from 1/2 grows with $n$, the number of repeated pulses.  (c) The calibration of the DRAG parameter performs the  $X_{\pi/2}-X_{-\pi/2}$ sequence while varying $\lambda$, the amplitude of the derivative pulse on the quadrature channel.  The correct derivative amplitude corresponds to the point where the qubit returns to the ground state.  }
	\label{fig:cals_example}
\end{figure}

We use single sideband~(SSB) modulation of our control pulses and calibrate the in-phase/quadrature (IQ) mixers (MITEQ IRM0408LC2Q) for the chosen intermodulation frequency (IF) to ensure only the correct sideband was produced with minimal leakage at the carrier frequency. We then calibrate the in-phase control pulse amplitude and the amplitude of the quadrature component for the DRAG correction.  The pulse amplitudes for a $\pi$-pulse ($X_{\pi}$) and a $\pi/2$-pulse ($X_{\pi/2}$) about the $x$-axis are tuned up by repeating the pulses in the sequence $X_{\pi/2} - (X_{\{\pi,\pi/2\}})^{2n}$ in order to amplify the errors.  The evolution of the qubit's Bloch vector during the first three points of this sequence is depicted in Fig. \ref{fig:cals_example}(a). 

We correct for over- or under-rotations by fitting to the measured population of the qubit ground state, $P(\ket{0})$ [see Fig. \ref{fig:cals_example}(b)].  Under the assumption that the error is only an over- or underrotation, it is simple to derive a fitting formula for the amplitude calibration sequences. The fit function for the $X_{\pi/2}$ pulse in this sequence is
\begin{equation}
P(\ket{0}) = a+\left(\frac{1}{2}(-1)^n \cos(\pi/2+2 n \epsilon)\right),
\label{eq:calFit}
\end{equation}
where $a$ is left as a fit parameter and goes to $1/2$ for perfect $X_{\pi/2}$ pulses.  For $X_{\pi}$ the fit function is
\begin{equation}
P(\ket{0}) = a+\left(\frac{1}{2} \cos(\pi/2+2 n \epsilon)\right).
\label{eq:calFitPi}
\end{equation}
The angle error, $\epsilon$, found by this fit corresponds to a gate error $r \approx \epsilon^2/6$. After fitting the error, we update the pulse amplitude accordingly.

Lastly, we determine the DRAG correction by applying the sequence $(X_{\pi/2}-X_{-\pi/2})$ while varying the amplitude of the derivative pulse on the quadrature channel [Fig. \ref{fig:cals_example}(c)].  The final state of the qubit traces a cosine as a function of this DRAG amplitude, and we select the value that returns the qubit in the ground state, $\lvert0\rangle$.   

The calibrated pulses are used for iterative randomized benchmarking (IRB), in which we interleave each target sequence zero to 16 times within random sequences of up to 365 Clifford gates [as depicted in Fig.~\ref{fig:tickPlot}(b)].  We average over 35 instances of each sequence and fit the decay to $A_n\alpha_n^i+B_n$, where $i$ is the number of Clifford gates, and $n$ is the number of interleaved gates.  Error bars are equal to the 95$\%$ confidence interval of this fit.

\begin{figure}
	\centering
		\includegraphics[width=0.8\columnwidth]{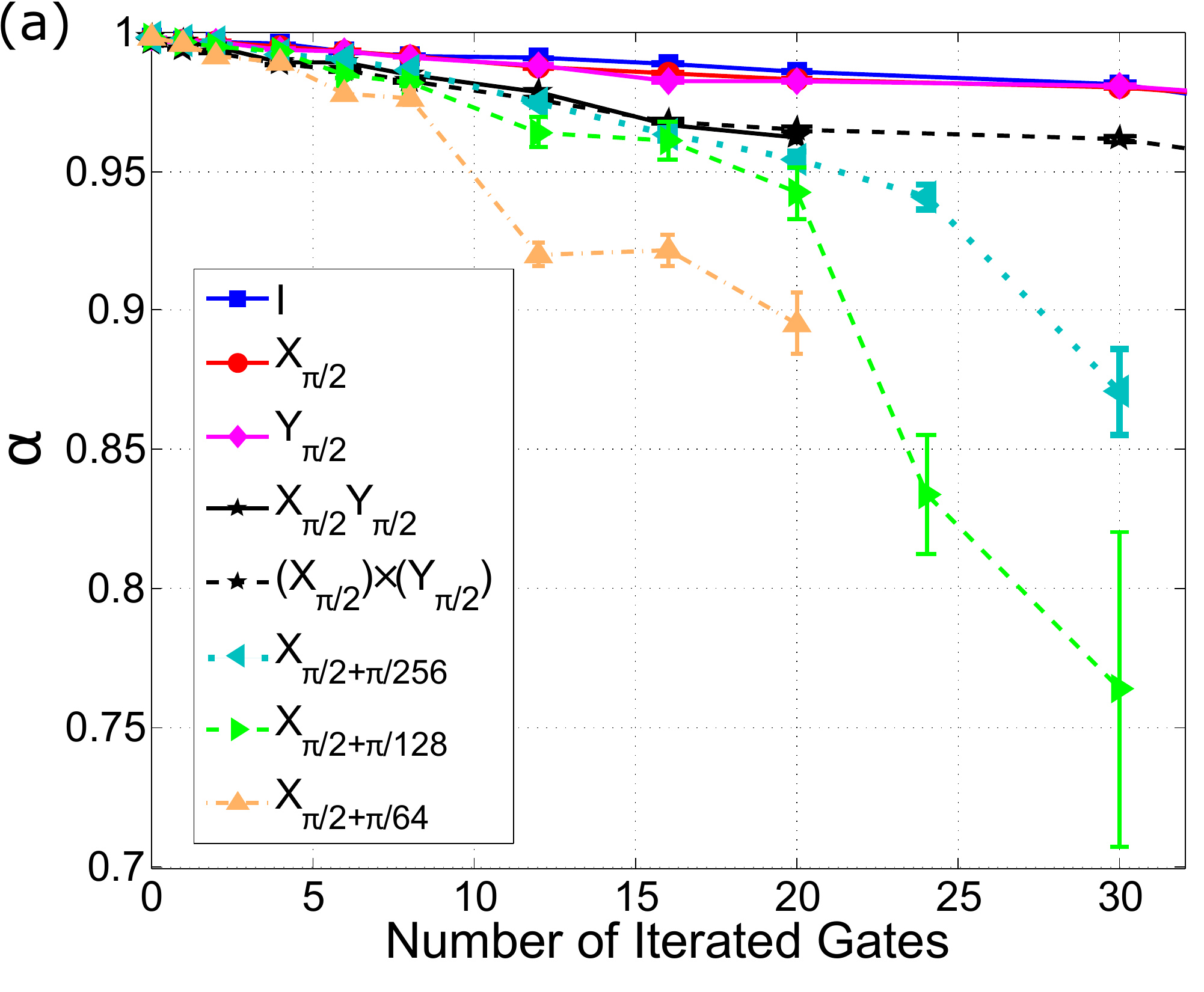}
		\includegraphics[width=0.8\columnwidth]{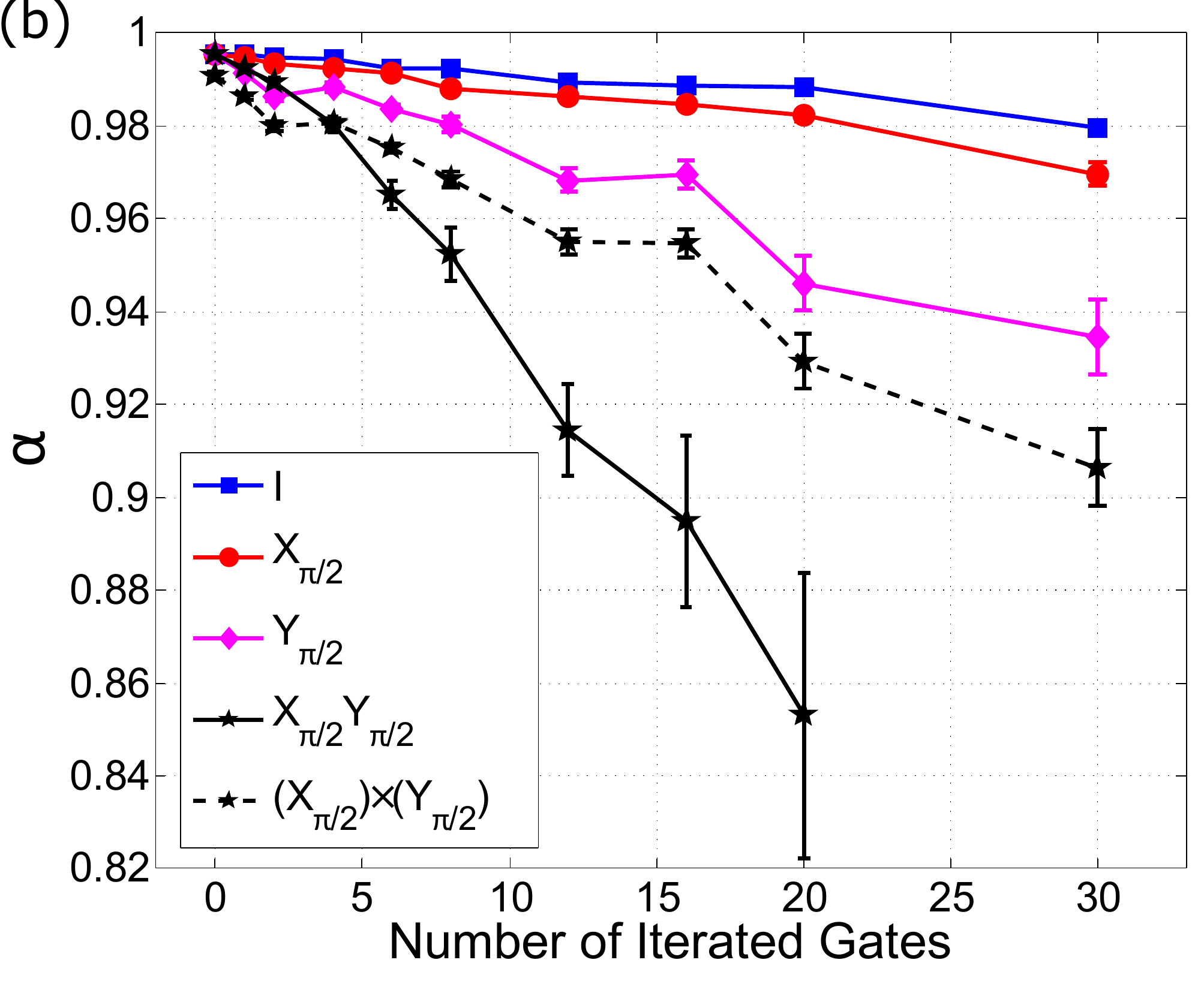}
	\caption{(color online) Iterative benchmarking data for (a)~a $16.7\,\text{ns}$ gate and (b)~a $10.0\,\text{ns}$ gate. The interleaved gates are the identity (blue squares), $X_{\pi/2}$ (red circles), $Y_{\pi/2}$ (magenta diamonds), and $X_{\pi/2} Y_{\pi/2}$ (black stars).  The product of $\alpha$ for $X_{\pi/2}$ and $Y_{\pi/2}$ is shown (dashed black stars) for comparison to the $X_{\pi/2} Y_{\pi/2}$ gate.  Also in (a) are interleaved overrotations on an $X_{\pi/2}$ by $\pi/256$ (aqua triangles), $\pi/128$ (dotted green triangles), and $\pi/64$ (dashed orange triangles). The error bars here are the 95$\%$ confidence interval of the fit to the IRB data averaged over 35 instances.}
	\label{fig:iterativeRB}
\end{figure}

We performed this protocol with a $16.7\,\text{ns}$ gate time [the time producing the minimum error per gate, Fig.~\ref{fig:tickPlot}(a)] and interleave the targets $I$, $X_\pi$, and $X_{\pi/2}$.  For these three gates, the decay in $\alpha$ versus the number interleaved gates is linear [Fig.~\ref{fig:iterativeRB}(a)].
This is consistent with the RB data that suggests the unitary errors at this gate time are small.

We then intentionally add overrotation errors to the $X_{\pi}$ gate to determine a bound on the sensitivity of this procedure to amplitude errors.  
We repeat the iterative benchmarking procedure with the $X_{\pi/2}$ pulse replaced with $X_{\pi/2+\epsilon}$, where $\epsilon = \{\pi/64$, $\pi/128$, $\pi/256\}$.
The $\pi/64$ and $\pi/128$ overrotations lead to fidelities that fall off quadratically and are clearly distinguishable from gates approaching the coherence limit.  The $\pi/256$ appears to have similar errors to the calibrated gates, giving a bound on the sensitivity to overrotation errors. Note that with infinite $T_1$ we could increase the sensitivity of this scheme by repeating a larger number of interleaved gates.

In order to quantify the amount of unitary versus non-unitary errors in the iterative randomized benchmarking data, we fit the data to both quadratic and
linear models.  Using the Akaike information criterion~(AIC), we determine which model most accurately describes the data \cite{akaike,burnham}.  The AIC is a useful tool for model selection and has been applied to quantum information previously \cite{vanenk}.

For $n$ data points and $k$ fitting parameters, the AIC is given by
\begin{equation}
C = n \ln\Bigl(\frac{R}{n}\Bigr) +2k+\frac{2k(k+1)}{n-k-1},
\end{equation}
where $R$ is the residual sum of squares for the fit.   The final term in this expression is a correction under the condition that $n < 40k$. This correction increases the penalty for overfitting when the sample size is small. We compute the $C$ for three models: linear, quadratic with no linear component, and combined linear and quadratic (see Table \ref{tab:AICvalues}).  The relative probability that the $i$th model is correct is 
\begin{equation}
P_i = \exp\left[\frac{1}{2}\bigl(C_{\mathrm{min}}-C_i\bigr)\right],
\end{equation}
with $C_{\mathrm{min}}$ the smallest AIC value for the set of models.  The model with the best fit to the data will have $P_i = 1$.  We calculate the relative probabilities for the three models for iterative randomized benchmarking data with $X_{\pi/2}$ pulses with no overrotation, $\pi/128$ and $\pi/256$ overrotations. As detailed in Table \ref{tab:AICvalues}, the calibrated gate with no added error is best fit by a linear model, as expected when there is little unitary error present.  The gate with $\pi/256$ overrotation is fit best by the combined model.  The preferred model according to the AIC for the gate with $\pi/128$ error is the quadratic model, but this is in part due to the penalty placed on adding extra parameters to the fit function.

\begin{table}[tb]
\vspace{5mm}
	\centering
		\begin{tabular}{c|ccc}
			\hline\hline
			Fit Function&0&$\pi/256$&$\pi/128$\\
			\hline
			$ax+b$&1&$1.3\times 10^{-3}$&$2.2\times 10^{-3}$\\
			$ax^2+b$&$2.0\times 10^{-7}$&0.18&1\\
			$ax^2+bx+c$&0.29&1&0.16\\
			\hline\hline
		\end{tabular}
	\caption{AIC values for gates with no overrotation, $\pi/256$ overrotation, and $\pi/128$ overrotation for linear and quadratic model functions.}
	\label{tab:AICvalues}
\end{table}

From this analysis it follows that a $\pi/128$ overrotation is detectable with this method and that consequently coherent rotation errors must be smaller than this value. We therefore simulate RB in the presence of a systematic $\pi/64$ overrotation  (easily detectable by IRB were it present), demonstrating that this is not sufficient to explain the deviation of the experiment from the simulated RB [dashed black in Fig. \ref{fig:tickPlot}(a)]. We conclude that there is an additional source  of decoherence that is present under the continuous-driving conditions of an RB experiment. One possible form for such non-unitary error, would be a dephasing proportional to the Rabi rate of the drive, as would result from amplitude fluctuations in the local oscillator, an amplifier, or other microwave electronics along the control line. Simulated RB in presence of such noise (green dot-dashed) shows reasonable agreement with the experimental data.  Drive noise with a $1/f$ dependence has been measured in flux qubits \cite{yoshihara}, and such low freqeuncy noise has been studied in the context of randomized benchmarking \cite{fogarty,epstein}.

We notice that there is still a deviation from the best fit at the shortest gate time in Fig. \ref{fig:tickPlot}(a).   To understand the origin of this larger error rate we calibrate gates of length $10\,\text{ns}$ and apply IRB.  For interleaved $I$, $X_{\pi/2}$, and $Y_{\pi/2}$ the iterative benchmarking data appears to decay linearly [Fig.~\ref{fig:iterativeRB}(b)].  First, we notice that the error of a $Y_{\pi/2}$ gate is larger than the $X_\pi/2$ gate error. We attribute this to our calibration procedure, in which the amplitude of the $Y_{\pi/2}$ is assumed to be equal to the $X_{\pi/2}$ pulse amplitude, but sampling errors in the pulse generation are not taken into account. Second, when the interleaved sequence is $X_{\pi/2} Y_{\pi/2}$ (black stars) a larger decay is observed.  This cannot be accounted for by multiplying (dashed black stars) the individual benchmarking parameters, $\alpha$, for the $X_{\pi/2}$ (red circles) and $Y_{\pi/2}$ (magenta diamonds) implying an additional error on the $X_{\pi/2} Y_{\pi/2}$ gate. (Note that, in contrast, no additional error for the $X_{\pi/2} Y_{\pi/2}$ sequence is observed for the 16.67 ns gate, for which the product of $X_{\pi/2}$ and $Y_{\pi/2}$ matches the error for $X_{\pi/2} Y_{\pi/2}$.)
The $X_{\pi/2}Y_{\pi/2}$ is not directly calibrated, and the presence of unitary errors here indicates a phase error, despite the fact that SSB modulation ensures the orthogonality of X and Y pulses by imposing a $\pi/2$ phase shift on the IF signal. 

After identifying the phase error, we have developed an error amplification sequence similar to those of Fig. \ref{fig:cals_example} in order to quantify an X-Y axes error.  The sequence is a repetition of $X_\pi Y_\pi$ within a Ramsey experiment:
$$X_{\pi/2} - \left(X_{\pi} - Y_{\pi} \right)^n - Y_{-\pi/2}.$$
The fit function for the error case when $X$ and $Y$ are not orthogonal is the same function as for a $\pi/2$ amplitude error given in Eq. \ref{eq:calFit}. The gate error measured by this sequence is $2\epsilon^2/3$.

\begin{figure}
	\centering
	\vspace{5mm}
		\includegraphics[width = \columnwidth]{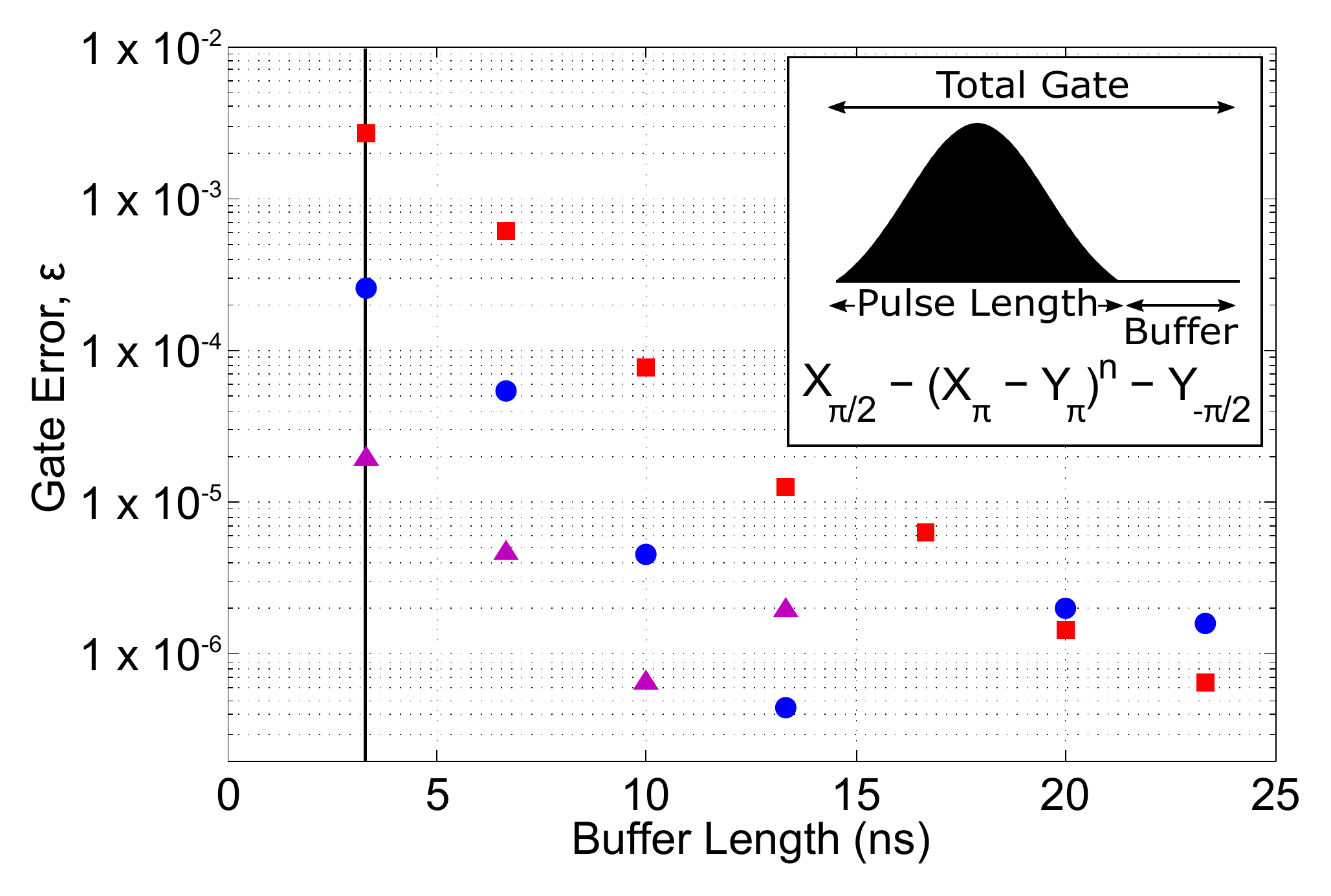}
		\caption{The gate error measured as a fit to the error amplification sequence $X_{\pi/2} - \left(X_{\pi} - Y_{\pi} \right)^n - Y_{\pi/2}$.  The gate error is plotted versus buffer length for three pulse lengths: $6.67\,\text{ns}$ in red squares, $10\,\text{ns}$ in blue circles, and $13.33\,\text{ns}$ in violet triangles.  The buffer length used for the data taken in Fig. \ref{fig:iterativeRB} was the shortest one shown here, $3.33\,\text{ns}$ (indicated by the solid vertical line).}
	\label{fig:bufferVSgateLength}
\end{figure}

We measure this error as a function of the buffer time between pulses for three different pulse lengths, as shown in Fig. \ref{fig:bufferVSgateLength}.  The IRB data was taken with a 3.33 ns buffer indicated by the vertical line [with pulse length of 13.33 ns for the data in Fig. \ref{fig:iterativeRB}(a) and 6.67ns for Fig. \ref{fig:iterativeRB}(b)].  The gate error is $2\times 10^{-5}$ for the pulse length corresponding to the 16.67 ns gate, and $3\times 10^{-3}$ for the 10 ns gate. This is consistent with the IRB data that demonstrates an axis error is present for the 6.67 ns pulse (red squares in Fig. \ref{fig:bufferVSgateLength}) but is not detected for 13.33 ns (violet triangles).   The gate error decreases as the buffer time is increased until it levels off by around 15 ns, at which point the resolution of the fit is not better than $1\times 10^{-5}$.   Because the error decreases with longer buffer time, it is likely due to distortions that cause successive pulses to overlap when the time between them is insufficient.  Note that this effect is not typically considered in RB, in which it is assumed a pulse knows no history of previous pulses in the sequence.  This pulse distortion may be alleviated by further pulse shaping (as shown in \cite{gustavsson} with pulse distortions on flux qubits) and will be the subject of future investigations.

We have introduced a variation of randomized benchmarking, useful for distinguishing non-unitary from unitary errors, and have validated this method on a superconducting qubit experiment. IRB will work for most physical unitaries without knowledge of the type of error present.  Once a unitary error is discovered, one can develop a calibration sequence to reduce the error.  By pushing gate lengths down and paying careful attention to calibrating the resulting unitary errors, we have achieved a benchmarked single-qubit gate fidelity of $99.95\%$. The error rate corresponding to this fidelity still deviates from the expected coherence by about a factor of four, but our iterative randomized benchmarking data indicates that we are \textit{not} limited by unitary errors at this point. We now seek to identify sources of drive-activated non-unitary errors (beyond $T_1$ and $T_2$) that must be limiting our fidelity at this time.  

We acknowledge discussions and contributions from Oliver E. Dial, Matthias Steffen, George A. Keefe, and Mary B. Rothwell.  This work was supported by ARO under contract W911NF-14-1-0124. 

\bibliography{iterativeBM}

\end{document}